\documentclass[aps,twocolumn,notitlepage,footinbib,amsmath,amssymb,amsfonts,superscriptaddress]{revtex4-1}

\usepackage{graphicx}
\usepackage{multirow}
\usepackage{color}
\usepackage{enumerate}
\usepackage[normalem]{ulem}
\newcommand{\mc}[1]{\ensuremath{\mathcal{#1}}}
\newcommand{\mr}[1]{\ensuremath{\mathrm{#1}}}

\newcommand{\bra}[1]{\ensuremath{\langle #1 |}}
\newcommand{\ket}[1]{\ensuremath{| #1 \rangle}}

\newcommand{\ua}{\ensuremath{{\uparrow}}}
\newcommand{\da}{\ensuremath{{\downarrow}}}

\newcommand{\bath}{\mathrm{bath}}
\newcommand{\Tk}{T_\mathrm{K}}
\newcommand{\PK}{P_\mathrm{K}}
\newcommand{\PT}{P_\mathrm{T}}
\newcommand{\oneCRK}{{\scriptscriptstyle{\mathrm{1CRK}}}}
\newcommand{\twoCRK}{{\scriptscriptstyle{\mathrm{2CRK}}}}

\newcommand{\Simp}{S_{\mathrm{imp}}}
\newcommand{\omegamax}{\omega_{\mathrm{max}}}

\newcommand{\etaone}{\eta_1}
\newcommand{\etatwo}{\eta_2}

\newcommand{\oneck}{{\scriptscriptstyle \textrm{1CK}}}
\newcommand{\twock}{{\scriptscriptstyle \textrm{2CK}}}
\newcommand{\onecark}{{\scriptscriptstyle \textrm{1CARK}}}
\newcommand{\twocark}{{\scriptscriptstyle \textrm{2CARK}}}
\newcommand{\onecsrk}{{\scriptscriptstyle \textrm{1CSRK}}}
\newcommand{\twocsrk}{{\scriptscriptstyle \textrm{2CSRK}}}

\newcommand{\Fig}[1]{Fig.~\ref{#1}}

\newcommand{\Sec}[1]{Sec.~\ref{#1}}
\newcommand{\App}[1]{App.~\ref{#1}}
\newcommand{\Ref}[1]{Ref.~\cite{#1}}

\allowdisplaybreaks

\RequirePackage[
   hyperindex,colorlinks,bookmarksnumbered,
  plainpages=true,pdfstartview=FitH]{hyperref}
 \hypersetup{linkcolor=blue,urlcolor=blue,citecolor=blue}
\usepackage{hyperref}

\begin{document}

\title{Non-Fermi-liquid Kondo screening under Rabi driving}
\author{Seung-Sup B. Lee}
\author{Jan von Delft}
\affiliation{Faculty of Physics, Arnold Sommerfeld Center for Theoretical Physics, Center for NanoScience, and Munich Center for Quantum Science and Technology,
Ludwig-Maximilians-Universit\"{a}t M\"{u}nchen, Theresienstra{\ss}e 37, 80333 M\"{u}nchen, Germany}
\author{Moshe Goldstein}
\affiliation{Raymond and Beverly Sackler School of Physics and Astronomy, Tel Aviv University, Tel Aviv 6997801, Israel}
\date{\today}
\begin{abstract}
We investigate a Rabi-Kondo model describing an optically driven
two-channel quantum dot device featuring a non-Fermi-liquid Kondo
effect.  Optically induced Rabi oscillation between the valence and
conduction levels of the dot gives rise to a two-stage Kondo effect:
Primary screening of the local spin is followed by secondary nonequilibrium
screening of the local orbital degree of freedom. Using bosonization arguments and the
numerical renormalization group, we compute the dot emission
spectrum and residual entropy. Remarkably, both exhibit two-stage Kondo
screening with non-Fermi-liquid properties at both stages.
\end{abstract}
\maketitle

\section{Introduction}
\label{sec:introduction}

The Kondo effect, involving a local spin entangled with a bath of
delocalized electrons, has been studied extensively in bulk systems
and in transport through quantum dots. Some years ago, a landmark
experiment~\cite{Latta2011} showed that it can also be probed
optically: A weakly driven optical transition between
the valence and conduction levels of the dot was used to abruptly switch the Kondo
effect on or off, leaving telltale power-law signatures~\cite{Tuereci2011} in the dot emission spectrum.  The case of strong  
spin-selective optical driving was subsequently studied theoretically within the
context of a single-channel Rabi-Kondo (1CRK) model~\cite{Sbierski2013},
involving Rabi oscillations between the dot valence and conduction
levels. This was predicted to lead to a novel nonequilibrium
quantum-correlated state featuring two-stage Kondo screening: The
local spin is screened by a primary screening cloud via the single-channel Kondo (1CK) effect, then
the Rabi-driven levels by a larger, secondary screening cloud.
Despite its nonequilibrium nature, this state has
a simple Fermi-liquid (FL) description in terms of scattering
phase shifts, since only a single screening channel is involved. 
   
This raises an intriguing question: What type of nonequilibrium state will
arise when the Rabi-driven dot couples to \textit{two} spinful channels,
described by a two-channel Rabi-Kondo (2CRK) model?
Without Rabi driving, it reduces to the standard two-channel Kondo
(2CK) model, known to have a non-Fermi liquid (NFL) ground state
\cite{Nozieres1980}, describable by Bethe Ansatz 
\cite{Tsvelick1985,Tsvelick1985a,Andrei1984}, conformal field theory (CFT)
\cite{Affleck1991b,Affleck1991,Affleck1993} or bosonization   
\cite{Emery1992,Emery1993,vonDelft1998,Zarand2000}.
However, NFL physics is known to be very sensitive to perturbations such as channel asymmetry or a magnetic field.
Do the NFL
properties survive under Rabi driving? If so, what are their fingerprints? In this paper, we answer these questions. We use a combination of bosonization arguments and numerical renormalization group (NRG) 
\cite{Wilson1975,Bulla2008,Weichselbaum2012:mps} calculations to compute the 2CRK emission
spectrum and impurity entropy. We find that NFL behavior survives, and, remarkably, leaves clear fingerprints in the emission in both the primary \textit{and} secondary screening regimes.

The rest of this paper is organized as follows.
In \Sec{sec:2CRK}, we introduce our system, the 2CRK model.
In \Sec{sec:qualtitative}, we provide a qualitative description of the screening processes in the 2CRK model.
In Secs.~\ref{sec:entropy} and \ref{sec:cloud}, we study the impurity contribution to the entropy and the Kondo cloud, respectively.
In \Sec{sec:bosonization}, the main points of our bosonization approach are outlined.
In \Sec{sec:spectrum}, we analyze the emission spectrum.
We conclude in \Sec{sec:conclusions}.
\App{app:bosonization} offers the details of our bosonization approach.

\section{Two-channel Rabi-Kondo model}
\label{sec:2CRK}

In this section, we first introduce the system in the lab frame, and then derive the effective Hamiltonian in the rotating frame to be treated by NRG and bosonization.

\begin{figure}
\centerline{\includegraphics[width=0.98\linewidth]{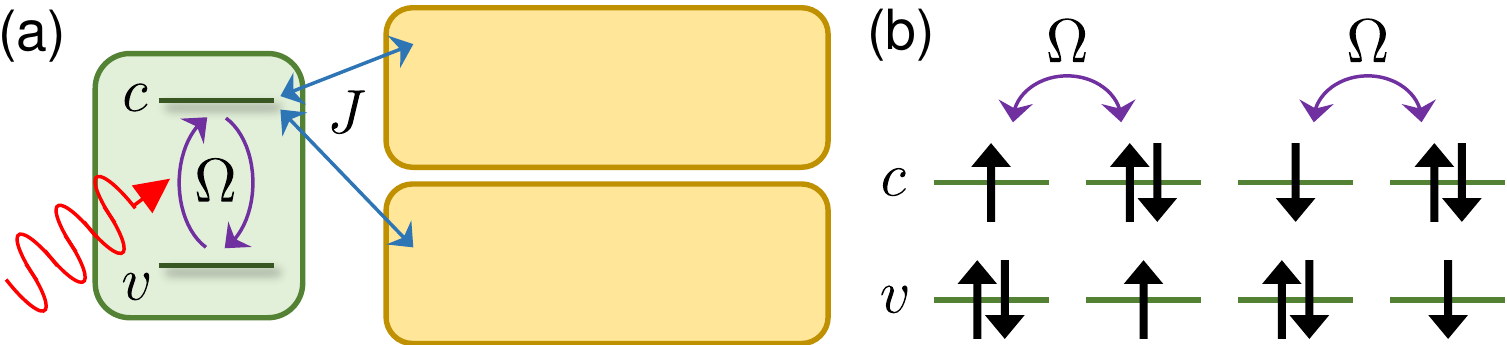}}
\caption{%
(a) Schematic depiction of the 2CRK model in the lab frame.
A small dot with two levels (conduction $c$ and valence $v$) has its $c$ level coupled to two large dots via spin exchange $J$.
The analogous 1CRK model has only one large dot.
Linearly polarized laser induces Rabi oscillation of frequency $\Omega$ in which electrons transition between the $c$ and $v$ levels, accompanied by the absorption and emission of light.
(b) The states of the small dot having $n_d = n_c + n_v = 3$ electrons.
The states of $(n_c, n_v) = (2,1)$ are connected to the states of $(n_c, n_v) = (1,2)$ via the Rabi oscillation.
In the rotating frame, the $c$-$v$ coupling becomes time-independent with amplitude $\Omega$.
}
\label{fig:setup}
\end{figure}

We consider a small quantum dot ($d$) with a conduction ($c$) and a valence ($v$) level as the impurity, and two large dots as the bath [\Fig{fig:setup}(a)].
The small dot is modelled by the Hamiltonian
\begin{equation}
H_d = \sum_{x = c, v} \left[ \frac{U_{xx}}{2} n_x (n_x - 1) + \epsilon_x n_x \right] + U_{cv} n_c n_v ,
\end{equation}
where $n_x = \sum_\sigma d_{x\sigma}^\dagger d_{x\sigma}$ denotes the particle number operator for the $x$ level ($x = c,v$), and $d_{x\sigma}$ annihilates spin-$\sigma$ electron at the $x$ level of energy $\epsilon_x$.
$U_{cc}$, $U_{vv}$, and $U_{cv}$ are the Coulomb interaction strengths.
The level separation $\epsilon_c - \epsilon_v$ is of the order of the semiconductor band gap $\sim 1 \, \mr{eV}$.
We consider the parameter regime in which the ground states of the small dot have $(n_c, n_v) = (1,2)$ in the absence of the Rabi driving to be introduced next.

We introduce a laser applied to the small dot, which induces Rabi oscillation between the $c$ and $v$ levels.
A circularly polarized laser would have coupled to one spin species due to an optical selection rule~\cite{Sbierski2013,Imamoglu1999}.
In the following we will consider the case of linearly polarized light, which symmetrically couples to both spin states.
The laser frequency $\omega_L$ is chosen to be close to the bare dot transition between $(n_c, n_v) = (1,2)$ states and $(n_c, n_v) = (2,1)$ states, i.e., $\omega_L \simeq U_{cc} + \epsilon_c - U_{vv} - \epsilon_v$.
(We set $\hbar = k_\mr{B} = 1$.)
Hence the $(n_c, n_v) = (2,1)$ states are accessed via the Rabi oscillation from the $(n_c, n_v) = (1,2)$ states [see \Fig{fig:setup}(b)].
The other states of $n_d = n_c + n_v \neq 3$ can be accessed only via virtual processes due to the energy cost of the Coulomb interaction.

Since the optical transition is close to the material's bandgap, that is, of order 1 eV, and much larger than all the other energy scales (which are typically not more than a few tens of meV), one could make the rotating wave approximation, under which a transfer of electron from the $v$ to the $c$ level involves the absorption of a photon and vice versa. We will further assume that the laser can be described as a classical field, and hence that spontaneous emission could be neglected.
Then the light-induced Hamiltonian term in the lab frame is given by
\begin{equation}
H_L^{\mr{(lab)}} = \Omega \sum_\sigma \left( d_{c\sigma}^\dagger d_{v\sigma} e^{- i \omega_L t} + \mr{h.c.} \right),
\end{equation}
where $\Omega$ is the Rabi frequency.

In addition, the $c$ level of the small dot is symmetrically tunnel-coupled to two identical large dots (channels $\ell = 1,2$).
These are assumed large enough to have essentially continuous excitation spectra, yet small enough that their charging energies suppress inter-channel charge transfer.
That is, $n_d + N_1$ and $n_d + N_2$ do not fluctuate, where $N_{\ell}$ means the particle number at the large dot $\ell$.
Under these conditions, the whole system Hamiltonian in the lab frame can be approximated, via the Schrieffer-Wolff transformation~\cite{Oreg2003} and up to an overall constant, by
\begin{equation}
\begin{aligned}
H^{\mr{(lab)}} &= \sum_\ell J \vec{S}_c \cdot \vec{s}_\ell + \delta_L n_v + H_\bath \\
&\quad + \Omega \sum_\sigma \left( d_{c\sigma}^\dagger d_{v\sigma} e^{- i \omega_L t} + \mr{h.c.} \right),
\end{aligned}
\end{equation}
where the Hilbert space for the small dot is restricted to the four-dimensional subspace of $n_d = 3$ shown in \Fig{fig:setup}(b).
Here $\vec{S}_c = \sum_{\sigma \sigma'} d_{c\sigma}^\dagger \tfrac{1}{2}
\vec{\sigma}_{\sigma \sigma'} d_{c\sigma'}$
and $\vec{s}_\ell = \sum_{\sigma \sigma'} \int_{-D}^{D} \mr{d}\epsilon \,
\mr{d}\epsilon' \tfrac{1}{2D} c_{\epsilon\ell\sigma}^\dagger \tfrac{1}{2}
\vec{\sigma}_{\sigma \sigma'} c_{\epsilon'\ell\sigma'}$
are $c$-level and $\ell$-channel spin operators, respectively.
$H_\bath = \sum_{\ell \sigma} \int_{-D}^D \mr{d}\epsilon \, \epsilon \, c_{\epsilon\ell\sigma}^\dagger c_{\epsilon\ell\sigma}$ describes the large dots with half-bandwidth $D$,
and $c_{\epsilon\ell\sigma}$ annihilates channel-$\ell$ electron of energy $\epsilon$ and spin $\sigma$.
The coupling strength $J$ is proportional to $1/(U_{cc} + 2 U_{cv} + \epsilon_c) - 1/(2 U_{cv} + \epsilon_c)$.

We will now go to the the rotating frame with respect to the laser-mode Hamiltonian, via the transformation $\mc{U}= e^{i \omega_L n_v t}$.
The rotating-frame Hamiltonian  $H^\mr{(rot)} = \mc{U}^\dagger H^\mr{(lab)} \mc{U} + i (\mr{d}\mc{U}^\dagger / \mr{d}t) \mc{U}$ will become time-independent,
\begin{equation}
\begin{aligned}
H^{\mr{(rot)}} &= \sum_\ell J \vec{S}_c \cdot \vec{s}_\ell + \delta_L n_v + H_\bath \\
& \quad + \Omega \sum_\sigma \left( d_{c\sigma}^\dagger d_{v\sigma}  + \mr{h.c.} \right),
\end{aligned}
\end{equation}
where $\delta_L = \omega_L - (U_{cc} + \epsilon_c - U_{vv} - \epsilon_v)$ is the detuning of laser frequency from the bare dot transition.
This is the 2CRK Hamiltonian to be studied in the rest of this paper.
For reference, we also include some results for the analogous 1CRK model ($\ell = 1$ only), and the standard 2CK and 1CK models (without $v$ level).

Since the coupling to the fermionic bath is assumed to be the main relaxation mechanism and dominates over spontaneous emission, the system would relax to an electronic equilibrium state in the rotating frame, which corresponds to a time-dependent state in the lab frame.
Thus we can analyze the system in the rotating frame employing equilibrium concepts such as entropy.

Note that our setup, which is driven optically, is different from previous setups driven by ac magnetic field~\cite{Zvyagin2005,Zvyagin2009}, in two key aspects.
First, the laser can be focused within the length scale of optical wavelength, so one can selectively drive the small dot only.
This selectivity has been demonstrated in experiments~\cite{Latta2011}.
Second, the rotating wave approximation works very well for our system, since the energy scale of the laser frequency is larger than the other energy scales in the system by at least two orders of magnitudes.
The selectivity and the rotating wave approximation are, however, unlikely for the systems driven by ac magnetic field that are in the microwave or rf regime.

\section{Qualitative considerations}
\label{sec:qualtitative}

Without Rabi driving, $\Omega = 0$, the ``trion'' and ``Kondo'' sectors, 
with $c$ and $v$ level occupancies $(n_c, n_v) = (2,1)$ and
$(1,2)$, respectively, are decoupled, and the $v$ level is 
inert.  The trion sector is a trivial FL, with the doubly occupied
$c$ level forming a local spin singlet. The Kondo sector
constitutes a standard Kondo model, involving the spin of the
singly-occupied $c$ level.  Below a characteristic Kondo temperature
$\Tk$, it will be screened by bath electrons. For the 2CRK
model, it is overscreened, leading to NFL behavior characteristic of
the 2CK model.  For the 1CRK model, it is fully screened, showing 
standard 1CK FL behavior.

For \textit{weak driving}, 
$0 < \Omega \ll \Tk$, Rabi oscillations
between the $c$ and $v$ levels couple the Kondo and trion sectors. 
Then primary screening of the
$c$-level spin, occurring at energies $\lesssim \Tk$, will be followed
by secondary screening of $c$-$v$ transitions at the renormalized Rabi coupling $\Omega^*$
(as in Ref.~\cite{Sbierski2013}),
provided that the ground state energies of the two (decoupled) sectors
differ by less than $\Omega^*$.
(A precise definition of $\Omega^*$ will be given later.)
We thus fine-tune $\delta_L$ such that for $\Omega=0$ the Kondo and trion ground states are
degenerate, following a strategy discussed in the Supplemental Fig. S2 of \Ref{Sbierski2013}.

Finally, for \textit{strong driving}, $\Omega \gtrsim \Tk$, 
the Rabi coupling generates
a strong splitting of bonding and anti-bonding states built from the
$c$ and $v$ levels. The local spin of the bonding state
will then undergo single-stage screening, as for the 
standard 2CK or 1CK models. 

These qualitative arguments will be substantiated quantitatively below by NRG calculations and bosonization arguments.
For the former,
we use $J=0.28 D$ throughout, leading to $\Tk^\twoCRK \simeq 3\times 10^{-4} D$ and $\Tk^\oneCRK \simeq 4\times 10^{-4} D$ when $\Omega=0$.
The bath discretization grid is set by $\Lambda^\twoCRK = 4$ and $\Lambda^\oneCRK = 2.7$,
and no $z$-averaging is used.
We use the QSpace tensor library~\cite{Weichselbaum2012:sym} to exploit the SU(2) symmetries of spin and channel where applicable.

\section{Entropy}
\label{sec:entropy}

\begin{figure}
	\centerline{\includegraphics[width=0.98\linewidth]{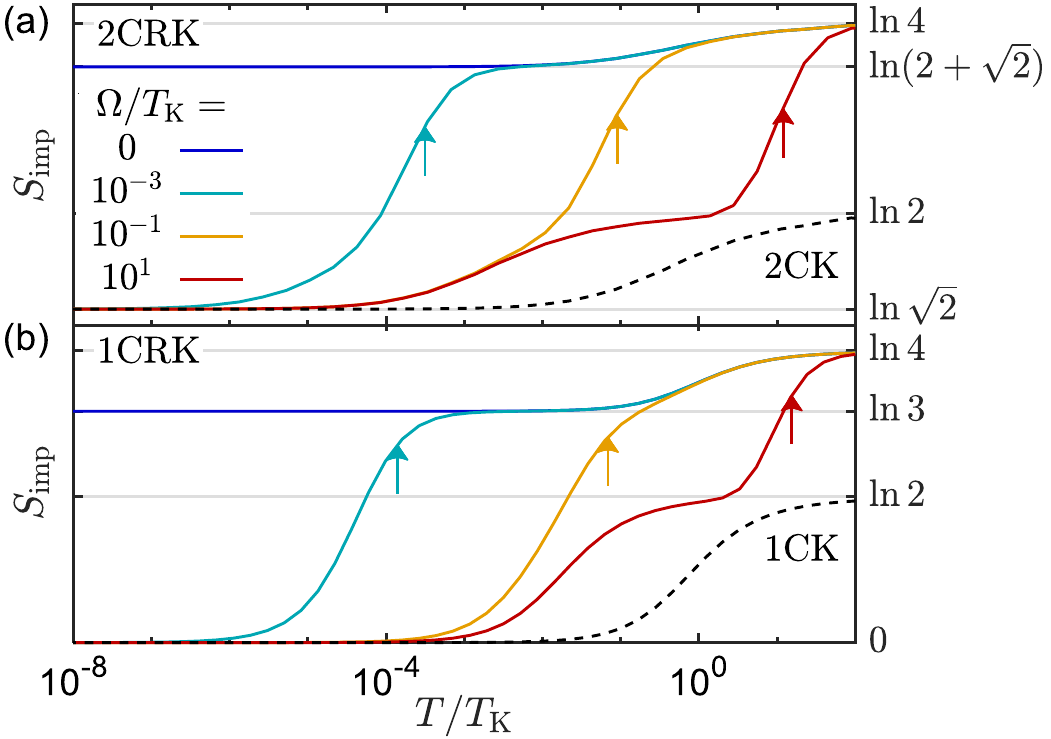}}
	\caption{Impurity contribution to the entropy, $\Simp$, for the (a) 2CRK and (b) 1CRK models, for four $\Omega$-values (solid lines).
	Arrows indicate the corresponding values of $\omegamax/\Tk$, the energy scale associated with the peak in the emission spectrum shown in \Fig{fig:ES}.
	For comparison, dashed lines show $\Simp$ for the standard 2CK and 1CK models, respectively, for the same value of $J$.%
	}
	\label{fig:Simp}
\end{figure}

Figure~\ref{fig:Simp}(a,b) shows our NRG results for the impurity contribution to the entropy~\cite{Bulla2008},
$\Simp$, which quantifies the effective degrees of freedom of the dot at different temperatures.
At high temperatures, $T \gg \Tk, \Omega$, the entropy $\Simp = \ln 4$ simply counts all
four configurations of the dot [\Fig{fig:Simp}(b)] for both the 2CRK and 1CRK models.
At lower temperatures, the behavior of the entropy depends on the relation of $\Omega$ and $\Tk$.

For strong driving $\Omega \gtrsim \Tk$, 
only two bonding states with different spins are accessible for $T < \Omega$.
Hence $\Simp (T \lesssim \Omega)$ shows a plateau at $\ln (2)$, followed by  a single crossover to $T = 0$ value of $\frac{1}{2}\ln ( 2) = \ln(\sqrt2)$ or $\ln (1)=0$ for the 2CRK or 1CRK models, respectively.
These values are the same as in the standard 2CK or 1CK models~\cite{Tsvelick1985a,Andrei1984,Affleck1991b} (shown as dashed lines), respectively. 
They reflect overscreening of a local spin by two spinful channels (resulting in a decoupled local Majorana mode~\cite{Emery1992,Emery1993,vonDelft1998,Zarand2000}), or its complete screening by a single spinful channel~\cite{Nozieres1974} (resulting in a spin singlet), respectively. 

In contrast, for weak driving $0 < \Omega \ll \Tk$, two-stage screening occurs.
For intermediate temperatures $\Simp (\Omega^* \ll T \ll \Tk)$ shows a primary-screening plateau at
$\ln (2 + \sqrt{2})$ or $\ln (2 + 1)$ for the 2CRK or 1CRK models: the
NFL- or FL-screened local spin contributes $\sqrt 2$ or 1 to the local
degeneracy count, with another 2 from the two trion ($v$) states.
At the lowest temperatures, $T \ll \Omega^*$, the $c$-$v$ transitions lead to a secondary-screening limiting value of
$\Simp = \ln \sqrt{2}$ or $0$ for the 2CRK and 1CRK models,
respectively, as for the standard 2CK and 1CK models. Finally,
for $\Omega=0$ (i.e., $\omegamax=0$), the primary-screening
plateau in $\Simp$ persists down to $T=0$.

\section{Kondo clouds}
\label{sec:cloud}

\begin{figure}
	\centerline{\includegraphics[width=0.98\linewidth]{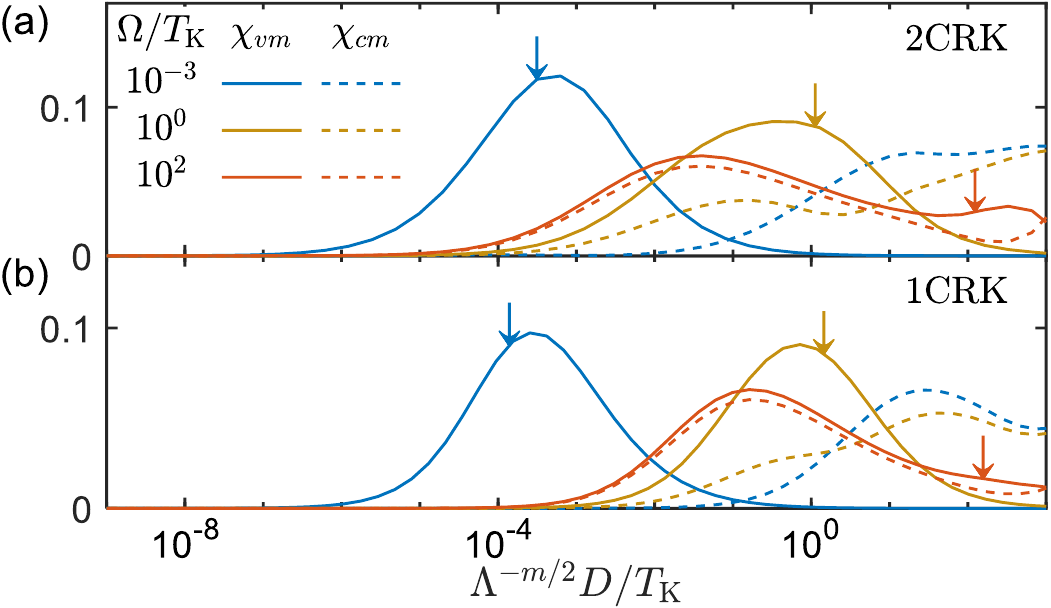}}
	\caption{Spin-spin correlators between the impurity and bath spin operators,
		revealing the structure of the screening clouds for the
		(a) 2CRK and (b) 1CRK models.
		We display $\chi_{vm} = -4 \langle \PT S_{vz} S_{mz} \rangle/\langle \PT \rangle$ (solid) and 
		$\chi_{cm} = -4 \langle \PK S_{cz} S_{mz} \rangle /\langle \PK \rangle$ (dashed),
		where $S_{vz}$, $S_{cz}$ and 
		$S_{mz}$ are $z$-component spin operators for the 
		$v$ level, $c$ level and the Wilson chain site $m \geq 0$, respectively.
		Site $m = 0$ is directly coupled to the $c$ level.
		$\PT = \sum_{\sigma} n_{v\sigma} (1 - n_{v\bar{\sigma}}) n_{c\ua} n_{c\da}$ and
		$\PK = \sum_{\sigma} n_{v\ua} n_{v\da} n_{c\sigma} (1 - n_{c\bar{\sigma}})$
		are projectors onto the trion and Kondo sectors,
		involving a singly-occupied $v$ or $c$ level, respectively.
		Both $\chi_{vm}$ and $\chi_{cm}$ are obtained by averaging two lines, interpolating odd and even $m$'s, respectively.
		We choose the abscissa as $\Lambda^{-m/2} D/\Tk$, where 
		$\Lambda^{-m/2} D$ is the energy scale (and also the inverse length scale~\cite{Wilson1975,Lee2015}) associated with the chain site $m$.
		For strong driving (red), $\chi_{cm}$ and $\chi_{vm}$ 
		have coinciding peaks, reflecting single-stage screening of
		the bonding-level spin. In contrast, for intermediate (yellow) and weak (blue) driving, we observe two-stage screening: 
		the peaks of $\chi_{cm}$, reflecting the screening of the $c$-level spin,
		occur at higher energies than those of
		$\chi_{vm}$, reflecting the screening of the $c$-$v$
		transitions. The area under each peak is $\simeq 1$. 
		Arrows indicate the corresponding values of $\omegamax/\Tk$.	
	}
	\label{fig:KC}
\end{figure}

To further study the nature of the screening clouds involved
in primary and secondary screening, 
we have computed spin-spin correlation functions between
the impurity and bath spin operators, see \Fig{fig:KC}. As described
in the caption thereof, for weak driving we find a nested, two-stage cloud,
screening the $c$-level spin at energies $\gtrsim \Tk$, and $c$-$v$ transitions at energies $\simeq \omegamax$.
In contrast, for strong driving we find just a single screening cloud.

\section{Bosonization}
\label{sec:bosonization}

We proceed to a more detailed analysis the weak driving case, $0 < \Omega \ll \Tk$, 
using bosonization (since the methods of \Ref{Sbierski2013} do not easily generalize to the 2CRK model).
Here we outline the main points, relegating further details to \App{app:bosonization}.
With uniaxial anisotropy, the bosonized form~\cite{Emery1992,Emery1993,vonDelft1998,Zarand2000} of 2CRK Hamiltonian $H_\mr{bath} + H_d$ is:
\begin{equation}
\begin{aligned}
H &= \sum_{\ell=1,2}
\left\{
\frac{u}{4\pi} \int_{-\infty}^\infty \mathrm{d}x \left[ \partial_x \phi_\ell(x) \right]^2 + 
\frac{J_z}{\pi\sqrt{2}} \PK S_z \partial_x \phi_\ell(0)
\right. \\
& \quad \left.
+ \frac{J_{xy}}{2\pi a} \PK
\bigl(S_+ e^{i\sqrt{2}\phi_\ell(0)} + \mathrm{h.c.}\bigr) \right\} + 2 \Omega \tau_x,
\end{aligned}
\end{equation}
where $S_\pm = S_x \pm i S_y$, while $\tau_+ = \sum_\sigma d_{c\sigma}^\dagger d_{v\sigma}$, $\tau_-=\tau_+^\dagger$, and $\tau_z = n_c-n_v$ are Pauli matrices in the orbital $c$-$v$ (Kondo-trion) pseudo-spin
space, and $\PK = (1+\tau_z)/2$ is
a projector onto the Kondo sector.
In addition, $u$ and $a = D/u$ are the Fermi velocity and lattice spacing (inverse momentum cutoff), 
and $\phi_\ell (x)$ is the chiral (unfolded) bosonic spin field (the charge sector decouples).
It obeys the commutation relation $[\phi_\ell (x),\phi_\ell (x')] = i \pi \, \mathrm{sgn} (x-x')$, where $\partial_x \phi_\ell (0)/(\pi\sqrt{2})$ is the density of the $z$-component of the channel-$\ell$ electron spin density at the dot site.

\subsection{1CRK}

Let us start from the single-channel case, where $\ell = 1$ [$\phi_2 (x)$ does not exist].
The unitary transformation $U_\alpha = e^{-i \alpha S_z \PK \phi_1(0)}$
with $\alpha =  J_z/(\pi \sqrt{2} u)$ eliminates the $J_z$ term at the cost of modifying the $J_{xy}$ term by a shift to the coefficient of $\phi_1(0)$ in the exponent.

At energies $\gg \Tk \gg \Omega$ we may ignore the Rabi term, and follow the usual perturbative renormalization group (RG) flow of the 1CK problem.
$J_{xy}$ flows since it has a nontrivial scaling dimension, set by the corresponding bosonic exponent (after the above-mentioned transformation). In addition, second-order spin-flip ($J_{xy}$) processes revive the non-spin-flip $J_z$ term, which may then be transformed away as above. $J_z$ thus flows to a fixed point value, $J_z=2\pi u$, corresponding to the Kondo fixed-point $\pi/2$ phase shift, while $J_{xy}$ grows until it becomes of the order of the reduced cutoff, which could serve to define the primary $c$-spin Kondo scale $\Tk$.
The $U_\alpha$-type transformations applied throughout the RG flow modify the Rabi term. Thus, below $\Tk$ we obtain the following intermediate-scale effective Hamiltonian:
\begin{equation}
\begin{aligned}
H_\mathrm{1CRK}^\mathrm{int} &=
\frac{u}{4\pi} \int_{-\infty}^\infty \mathrm{d}x \left[ \partial_x \phi_1 (x) \right]^2
+ \frac{J_{xy}^\mathrm{ren}}{\pi a} \PK S_x
\\ & \quad + \Omega  \tau_+ \left[ 
P_\uparrow e^{-i \phi_1 (0)/\sqrt{2}} + P_\downarrow e^{i\phi_1 (0)/\sqrt{2}} \right]  + \mathrm{h.c.},
\end{aligned}
\label{eqn:hi1cark_main}
\end{equation}
where $P_{\uparrow,\downarrow} = 1/2 \pm S_z$ is a projector into the subspace $S_z= \pm 1/2$, and $J_{xy}^\mathrm{ren} \sim \Tk \gg \Omega$. The latter large coupling fixes the dot spin to $S_x=1/2$, which corresponds, in the original basis, to an entangled state of the impurity and bath spins, i.e., the primary Kondo singlet. Thus $P_{\uparrow,\downarrow}$ are replaced by their expectation values $\langle P_{\uparrow,\downarrow} \rangle = 1/2$. 
The resulting model describes the hybridization between the pseudo-spin ($c$-$v$ or Kondo-trion) degree of freedom and the channel, which is equivalent (up to a transformation similar to $U_\alpha$ but involving $\tau_z$ instead of $S_z$) to an anisotropic Kondo model for the pseudo-spin space.
The Rabi coupling $\Omega$ is relevant, with scaling dimension $\eta_1 = 1/4$, determined by the corresponding bosonic exponent in Eq.~\eqref{eqn:hi1cark_main}, or, within CFT, from its role as boundary condition changing operator, turning on and off 1CK screening~\cite{Affleck1994}. Hence, $\Omega$ flows to strong coupling, creating a new scale, the renormalized Rabi frequency (secondary Kondo temperature), $\Omega^* \sim \Tk (\Omega/\Tk)^{1/(1-\eta_1)} = \Tk (\Omega/\Tk)^{4/3} \ll \Omega$, where one expects a peak in the dot emission spectrum to occur, instead of the more usual peak at $\Omega$ for strong driving $\Omega \gg \Tk$.
Below this scale, the pseudo-spin is screened by the creation of a secondary ``Kondo singlet''.

\subsection{2CRK}

Let us now perform a similar analysis of the 2CRK model.
Defining the fields $\phi_\pm(x)=[\phi_1(x) \pm \phi_2(x)]/\sqrt{2}$, only the former couples to $J_z$, and could be eliminated by a transformation similar to $U_{\alpha}$ defined with $\sqrt{2} \phi_+(0)$ instead of $\phi_1 (0)$.
For $\Omega \ll \Tk$ one may proceed with the primary 2CK RG flow, which drives $J_z$ to $\pi u$, corresponding to a $\pi/4$ phase shift, and $J_{xy}$ to $J_{xy}^\mathrm{ren} \propto \Tk \gg \Omega$. At the same time, the Rabi coupling gets modified. On the scale of $\Tk$ we thus arrive at:
\begin{equation}
\begin{aligned}
H_\mathrm{2CRK}^\mathrm{int} &=  \sum_{p=\pm} \frac{u}{4\pi} \int_{-\infty}^\infty 
\hspace{-.4em} \mathrm{d}x \left[ \partial_x \phi_p(x) \right]^2 +  
\frac{J_{xy}^\mathrm{ren}}{\pi a} \PK
S_x \cos \phi_{-}(0)
\\
& \quad
+ \Omega \tau_+ \left[ 
P_\uparrow e^{-i\phi_+(0)/2} + P_\downarrow e^{i\phi_+(0)/2} \right]  + \mathrm{h.c.} 
\end{aligned}
\label{eqn:hi2csrk_main}
\end{equation}
The first line describes the 2CK fixed point, at which the $\phi_-$ remains coupled: $S_x$ assumes a definite value $S_x=\pm 1/2$, and correspondingly $\phi_-(0)$ is locked to a minimum or maximum of the cosine function. Refermionizing the local spin-$\phi_-$ system, the $J_{xy}$ term couples a local Majorana fermion ($\propto S_x$) to the lead, leaving another local Majorana ($\propto S_y$) unscreened~\cite{Emery1992,Emery1993}.

We now turn to the second line. Since $J_{xy}^\mathrm{ren} \sim \Tk \gg \Omega$, we may again set $P_{\uparrow,\downarrow} \to 1/2$. The remaining term is a product of $\tau_\pm$ with bosonic exponents. The exponents contribute $1/8$ to the scaling dimension of $\Omega$, while $\tau_\pm$ turns on or off the $J_{xy}^\mathrm{ren}$ term, which is equivalent to turning on or off a local backscattering impurity in a Luttinger liquid, with scaling dimension 1/16~\cite{Gogolin1993,Prokofev1994}. Thus, the overall scaling dimension of $\Omega$ is $\eta_2=3/16$. This matches the corresponding CFT analysis of its role as a boundary-condition changing operator~\cite{Affleck1994}.
Thus, $\Omega$ is relevant, flowing to strong coupling and generating a new scale $\Omega^* \sim \Tk (\Omega/\Tk)^{1/(1-\eta_2)} = \Tk (\Omega/\Tk)^{16/13} \ll \Omega$, below which secondary screening of the $c$-$v$ (Kondo-trion) fluctuations is achieved. Importantly, since the Rabi term is spin symmetric, it does not interfere with the primary NFL 2CK screening, and leaves the decoupled Majorana ($S_y$) unscreened: While the Rabi term contains $S_z \propto S_x S_y$, the corresponding processes are suppressed by the dominant $J_{xy}^\mathrm{ren}$ term, and all higher order (in $\Omega$) processes which leave the system within the low-energy manifold of $J_{xy}^\mathrm{ren}$ term do not couple to $S_y$.

\begin{figure}
	\centerline{\includegraphics[width=0.98\linewidth]{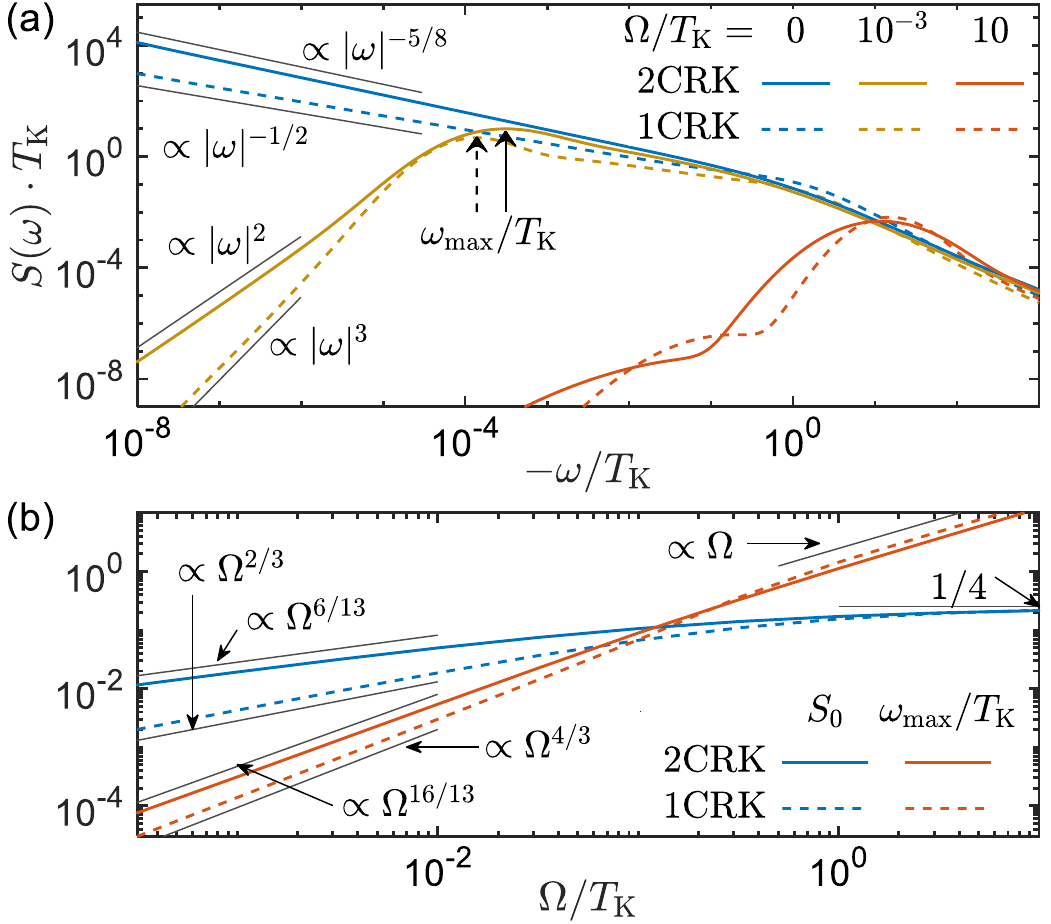}}
	\caption{
		(a) Log-log plot of the emission spectrum $S(\omega)$, and 
		(b) its finite-frequency peak position $\omegamax$ and zero-frequency spectral weight $S_0$
		as functions of Rabi driving $\Omega/\Tk$, for
		the 2CRK (solid) and 1CRK (dashed) models at $T = 0$.
		Guide-to-the-eye grey lines depict 
		the power laws predicted by bosonization arguments (see text).
	}
	\label{fig:ES}
\end{figure}

\section{Emission spectrum}
\label{sec:spectrum}

Having established the general picture of the two-stage NFL
screening, we can now analyze its effect on the main experimental observable,
the dot emission spectrum.
The emission spectrum of linear polarization at detuning $\omega$ from the driving laser frequency is proportional to the spectral function~\cite{Sbierski2013},
\begin{equation}
S (\omega) = \sum_{jj'} \rho_j \big| \bra{j'} {\textstyle \sum_\sigma} d_{v\sigma}^\dagger d_{c\sigma} \ket{j} \big|^2 \delta (\omega + E_{j'} - E_j) ,
\label{eq:S}
\vspace{-0.5\baselineskip}
\end{equation}
where $|j\rangle$ and $E_j$ are energy eigenstates and eigenvalues of the 
Rabi-Kondo Hamiltonian, and $\rho_j = e^{-E_j/T}/Z$. This is the spectral function of the Rabi term with itself.
At temperature $T=0$, the emission spectrum has weight only for $\omega \leq 0$.
Without Rabi driving, $S(\omega \to 0^-)$ shows a power-law divergence.
For weak driving, the divergence is cut off, giving way to a power-law decrease.
Accordingly a wide peak at $|\omega| = \omegamax$ and a delta-function peak $S_0 \delta(\omega)$ of weight $S_0$ at $\omega = 0$ emerge.
We identify $\omegamax$ with the renormalized Rabi frequency $\Omega^*$.

Figure~\ref{fig:ES}(a) shows a log-log plot of the emission spectrum, revealing its various power laws.
For weak driving, there are two distinct regimes: (i) The intermediate-detuning regime,
$\omegamax \lesssim |\omega| \lesssim \Tk$, is dominated by the Kondo
exchange coupling and reflects primary screening. Here the correlations of the Rabi term with itself are governed by its scaling dimension $\eta$, giving $S(\omega) \propto |\omega|^{2\eta-1}$ with $\eta = \eta_1 = 1/4$ and $\eta = \eta_2 = 3/16$ for the 1CRK or 2CRK models, respectively. Thus, this part of the spectrum reveals the scaling of 1CK vs.~2CK boundary condition changing operators~\cite{Affleck1994}.
(ii) The small-detuning regime, $|\omega| \lesssim \omegamax$, is dominated by the Rabi coupling and reflects secondary $c$-$v$ screening.
In this regime, $S(\omega)$ corresponds to the correlation function of the exchange interaction $\sum_\ell \vec{S}\cdot\vec{s}_\ell$ with itself in the standard Kondo models, which yields $S(\omega) \propto |\omega|^3$ and $\propto |\omega|^2$ for the 1CRK and 2CRK models, respectively.
The power is reduced in the 2CRK case, as the unscreened Majorana $S_y$ appearing in the Rabi term in Eq.~\eqref{eqn:hi2csrk_main} (through $S_z \propto S_x S_y$) reduces the corresponding scaling dimension by $1/2$.
Thus, the $|\omega|^{2}$-behavior is a clear fingerprint of the NFL nature of the nonequilibrium secondary screening nature in the 2CRK system.

Figure~\ref{fig:ES}(b) shows that $\omegamax$ and $S_0$ increase as power laws in $\Omega$. For weak driving, our previous analysis shows that, in accordance with the numerical data, $\omega_\mathrm{max} \sim \Omega^* \propto \Omega^{1/(1-\eta)} \sim \Omega^{16/13}$ or $\Omega^{4/3}$, and moreover (as we will momentarily explain), 
$S_0 \sim \Omega^{2\eta/(1-\eta)} \sim \Omega^{6/13}$ or $\Omega^{2/3}$ for the 2CRK or 1CRK
models, respectively.
Indeed, $S_0$ takes up the spectral weight missing at small detuning due to $\Omega^*$ cutting off the intermediate detuning $S(\omega) \propto |\omega|^{2\eta-1}$ behavior. Hence, $S_0 \propto \int_0^{\Omega^*} \mathrm{d}\omega \, |\omega|^{2\eta-1} \sim \Omega^{2\eta/(1-\eta)}$. Alternatively, by Eq.~\eqref{eq:S} $S_0$ is the square of the expectation value of $\tau_x$ (before transformations) in the ground state. But $\tau_x$ is the Rabi term divided by $\Omega$, and the Rabi term should scale as $\Omega^*$, leading to $S_0 \sim (\Omega^*/\Omega)^2 \sim \Omega^{2\eta/(1-\eta)}$, as before.
Thus, the Kondo boundary condition changing operators governs both $\omega_\mathrm{max}$ and $S_0$.
For strong driving, $\omegamax \sim \Omega$ corresponds to the transition energy between the bonding and anti-bonding states of $c$ and $v$ levels.

%\textit{Conclusions.---}%
\section{Conclusions}
\label{sec:conclusions}

We have identified a two-stage NFL screening process in a Rabi driven quantum dot.
The NFL nature survives in nonequilibrium as the Rabi driving respects both spin and channel symmetries.
We have developed a new bosonization approach that explains the power-law exponents obtained numerically.
The distinct power laws in the emission spectra should motivate optical spectroscopy studies on the multi-channel quantum dot devices.
The case of non-negligible spontaneous emission, which goes beyond the description of the time-independent Hamiltonian in the rotating frame, would be an interesting question for future study.
We envision our findings to also be relevant for higher-dimensional driven strongly-correlated materials.

\begin{acknowledgments}
We thank E.~Sela for useful discussions. This joint work was supported by German Israeli Foundation (Grant No.~I-1259-303.10).
S.-S.B.L.~and J.v.D.~are supported by the Deutsche Forschungsgemeinschaft (DFG, German Research Foundation) under Germany’s Excellence Strategy -- EXC-2111 -- 390814868;
S.-S.B.L.~further by Grant No.~LE 3883/2-1.
M.G.~acknowledges support by the Israel Science Foundation (Grant No.~227/15), the US-Israel Binational Science Foundation (Grant No.~2016224), and the Israel Ministry of Science and Technology (Contract No.~3-12419).
\end{acknowledgments}

% % % %
\appendix

\section{Bosonization details}
\label{app:bosonization}

In this Appendix, we develop in details the theory of the multichannel Kondo effect, by first reviewing the bosonization description of the ordinary single- and two-channel Kondo (1CK, 2CK), then going on to their Rabi-Kondo versions, without and with spin rotation symmetry (the latter being the case considered in the main text).

We note that the Yuval-Anderson (YA) Coulomb gas approach~\cite{Anderson1969,Yuval1970,Anderson1970,Vigman1978,Fabrizio1995a,Goldstein2009} is known to give equivalent results to the bosonization approach for all universal (i.e., cutoff-independent) quantities, such as critical dimensions.
Meanwhile, the Coulomb gas approach provides more accurate microscopic expressions for the phase shifts that are cutoff-dependent.
We have verified that the same is true for the systems discussed in this work.
However, in this paper we employ the bosonization approach, since it is more succinct than the Coulomb gas approach.

\subsection{Ordinary Kondo}

First, we review the ordinary (equilibrium) 1CK and 2CK effects from the bosonization perspective.

\subsubsection{Single-channel ordinary Kondo}

Let us start from the ordinary single-channel Kondo effect.  Using a
bosonic description of the channel, the charge sector decouples, while
the spin sector can be written in terms of a single right-moving
chiral boson over the entire 1D line (instead of a single non-chiral
boson on the 1D half-line), leading to the following Hamiltonian~\cite{Emery1992,Gogolin2004}:
\begin{equation}
\begin{aligned}
H_\oneck &= \frac{u}{4\pi} \int_{-\infty}^\infty \mathrm{d}x \left[ \partial_x \phi(x) \right]^2 + \frac{J_z}{\pi\sqrt{2}} S_z \partial_x \phi(0) \\
& \quad + \frac{J_{xy}}{2\pi a} \left(S_+ e^{i\sqrt{2}\phi(0)} + \mathrm{h.c.}\right) \, , 
\end{aligned}
\label{eqn:h1ck}  
\end{equation}     
where $S_\pm = S_x \pm i S_y$, $S_{x,y,z}$ are the impurity spin-1/2
operators, $u$ and $a$ are the Fermi velocity and lattice spacing
(inverse momentum cutoff), the bosonic field obeys the commutation
relation $[\phi(x),\phi(x^\prime)] = i \pi \mathrm{sgn} (x-x^\prime)$,
and $\partial_x \phi(0)/(\pi\sqrt{2})$ is the conduction electron spin density
at the dot site.
Applying the transformation
$H_\oneck \to H_\oneck^\prime = U_\alpha H_\oneck U_\alpha^\dagger$
where $U_\alpha = e^{-i \alpha S_z \phi(0)}$ with
$\alpha =  J_z/(\pi \sqrt{2} u)$, the $J_z$ term is eliminated, at the
cost of modifying the exponent in the $J_{xy}$ term:
\begin{equation}  
\begin{aligned}
H_\oneck^\prime &= \frac{u}{4\pi} \int_{-\infty}^\infty \mathrm{d}x \left[ \partial_x \phi(x) \right]^2 \\
& \quad + \frac{J_{xy}}{2\pi a} \left(S_+ e^{i\sqrt{2}[1-J_z/(2\pi v)]\phi(0)} + \mathrm{h.c.}\right) \, . 
\end{aligned}
\label{eqn:hp1ck}
\end{equation}
We now proceed with perturbative RG, using Cardy's operator product
expansion (OPE) version~\cite{Cardy1996}. $J_{xy}$ flows because it
has a nontrivial scaling dimension (due to the corresponding
nontrivial bosonic exponent), whereas the OPE of the two $J_{xy}$
terms reintroduces the $J_z$ term. This can be transformed again into
the bosonic exponent. Defining the dimensionless exchange couplings
$\mathcal{J}_{xy,z}=J_{xy,z}/(2\pi u)$, and denoting the energy cutoff
by $D=v/a$, we thus obtain Anderson's well-known RG equations:
\begin{align} \label{eqn:rgjxy}
  -D\frac{\mathrm{d} \mathcal{J}_{xy}}{\mathrm{d}D} & = \left[ 1 - \left(1-\mathcal{J}_z \right)^2 \right] \mathcal{J}_{xy}, \\
  -D\frac{\mathrm{d} \mathcal{J}_{z}}{\mathrm{d}D} & = \left(1-\mathcal{J}_z \right) \mathcal{J}_{xy}^2. \label{eqn:rgjz}
\end{align}
Thus, $\mathcal{J}_z$ flows to the strong-coupling fixed point value
$\mathcal{J}_z = 1$ ($\pi/2$ phase shift). At that point the impurity
spin becomes decoupled from the bath --- the exchange term becomes
simply $\propto \mathcal{J}_{xy} S_x$ (where at strong coupling
$\mathcal{J}_{xy} \propto \Tk$, the Kondo temperature), and seemingly
polarized the impurity spin in the $x$ direction. Recalling that the
$S_x$ operator has undergone a succession of transformations dressing
it with the bosonic field, we recognize that, in terms of the original
fields, this actually signifies (an anisotropic version of) the Kondo
singlet. 
Indeed, the fact that the spin flip terms in the original Hamiltonian,
$S_\pm e^{\pm i\sqrt{2}\phi(0)}$, have been renormalized to 
$S_\pm$ means that the renormalized versions of the
original $S_\pm$ operators are $S_\pm e^{\mp i\sqrt{2}\phi(0)}$.
The correlation function of these two operators decays in time as $1/t^2$ (due to the bosonic factor), in
accordance with Fermi liquid theory (in which one posits that at the
fixed point the impurity spin ``merges'' with the Fermi sea, so 
its correlator behaves like the correlation function of the lead
fermion density). Another way to get this result is to notice that,
generically (that is, in a higher order RG than what we considered),
$S_z$ could get dressed by the lead spin density at the impurity site,
$\propto \partial_x \phi(0)$, hence its correlation would decay as
$1/t^2$. Using similar arguments, the connected correlation function
of the exchange terms in the original Hamiltonian turns into a
connected correlator of two lead spin operators with two lead spin
operators, decaying as $1/t^4$, translating into an $\omega^3$
behavior of the corresponding spectral function at low frequencies.
Finally, since the impurity Hamiltonian reduces to $\propto \mathcal{J}_{xy} S_x$ at the fixed point, if a magnetic field in the $z$ direction is introduced, the
impurity susceptibility becomes
$\propto \mathcal{J}_{xy}^{-1} \propto \Tk^{-1} $. 
This will also give
a finite expectation value to the lead spin correlators, making the
leading contribution (at long time) to the exchange-exchange
correlation function decay as $1/t^2$, or $\omega$ in the frequency
domain.  Finally, the impurity entropy is $\ln 2$ at $T \gg \Tk$, and
goes to zero at $T \ll \Tk$, due to the Kondo screening.

\begin{widetext}
\subsubsection{Two-channel ordinary Kondo}

Now the starting Hamiltonian is:
\begin{equation} \label{eqn:h2ck}
  H_\twock = \sum_{\ell=1,2} \left\{ \frac{u}{4\pi} \int_{-\infty}^\infty \mathrm{d}x \left[ \partial_x \phi_\ell(x) \right]^2 + \frac{J_z}{\pi\sqrt{2}} S_z \partial_x \phi_\ell(0) + \frac{J_{xy}}{2\pi a} S_+ e^{i\sqrt{2}\phi_\ell(0)} + \mathrm{h.c.} \right\},
\end{equation}
where $\ell = 1,2$ labels the two conduction electron channels, $\partial_x \phi_\ell (0)/(\pi\sqrt{2})$
gives their respective spin densities at the dot site, 
and we assume channel symmetry. Here it is useful to define the symmetric and antisymmetric combinations, $\phi_\pm(x) = [\phi_L(x) \pm \phi_R(x)]/\sqrt{2}$, which keep the commutation relations the same. We now apply the transformation $U_{\alpha +} = e^{-i \alpha S_z \phi_+(0)}$ with $\alpha = J_z/(\pi u)$ to eliminate the $J_z$ term and get
\begin{equation} \label{eqn:hp2ck}
H_\twock = \sum_{p=\pm} \frac{u}{4\pi} \int_{-\infty}^\infty \mathrm{d}x \left[ \partial_x \phi_p(x) \right]^2 + \frac{J_{xy}}{\pi a} S_+ \cos\left[\phi_-(0)\right] 
\bigl( e^{i[1-J_z/(\pi v)]\phi_+(0) } + \mathrm{h.c.} \bigr) \, . 
\end{equation}
\end{widetext}

The RG equations are similar to the 1CK case, but with
$(1-\mathcal{J}_z) \to (1-2\mathcal{J}_z)$. Hence, $\mathcal{J}_z$
flows to a value of 1/2 ($\pi/4$ phase shift), at which point the
impurity remains coupled only to $\phi_-$. $J_{xy}$ continues to flow
to strong coupling, where it becomes $\propto \Tk$ (this strong
coupling bosonic description corresponds to the intermediate coupling
non-Fermi-liquid fixed point in the traditional description in terms
of the original fermions). If one refermionizes the local spin and the
bosonic subsystem $\phi_-$, the $J_{xy}$ term becomes a coupling of a
local Majorana operator ($S_x$) to a Majorana field density in the
lead at the adjacent site, namely $\cos[\phi_-(0)]$, while $S_y$
becomes a local decoupled Majorana. This is the famous Emery-Kivelson
point. Therefore, the low-temperature impurity entropy is
$\ln\sqrt{2}$.  Since $S_z \propto i S_x S_y$, its correlator with
itself is a convolution of the correlators of a localized
Majorana fermion ($\propto S_y$) and a propagating one ($\propto S_x$), and decays in time
as $1/t$, as that of one free fermion times one localized fermion,
leading to a logarithmic divergence of the susceptibility with the
largest cutoff energy (magnetic field, temperature, or frequency). For
a similar reason, the original non-spin-flip exchange term,
$\propto S_z \partial_x \phi_+(0)$, has correlations decaying as
$1/t^3$, implying a low-frequency power-law behavior of $\omega^2$, in the absence of a magnetic field 
(a magnetic field suppresses the non-Fermi-liquid 2CK physics, and
restores the Fermi-liquid 1CK $\omega$ behavior). If we look at
correlators of $S_+$ (with its conjugate), we can use the fact that
the series of transformations map it to $S_+ e^{-i\phi_+(0)}$,
leading to a $1/t$ behavior, similar to $S_z$.

One can recover the behavior of the susceptibility
using purely bosonic
language~\cite{Gogolin1994,Fabrizio1994,Fabrizio1995,Goldstein2010b}. At
the strong $J_{xy}$ fixed point, $S_x$ picks a value $\pm 1/2$, and
then $\phi_-(0)$ is pinned to either a minimum or a maximum of the
cosine function, respectively. With that one can calculate the
susceptibility, that is, the retarded correlator of $S_z$ with
itself. Indeed, $S_z$ anticommutes with $S_x$, hence with the
spin-flip exchange term.
Since the spin-flip exchange term modifies by unity the spin of one of
the leads, the operator $V = e^{i\pi N_-}$, where
$N_- = N_1 - N_2$ is the difference between the refermionized populations of the two
leads (corresponding to $S_z$ of the original electrons, since the
bosonic fields are all related to the original electronic spin degrees
of freedom), also anticommutes with the spin-flip exchange
term. Hence, the correlator of $S_z$ could be replaced by a correlator
of $V$. Conservation of the overall refermionized population,
$N_+ = N_1 - N_2$ allows one to write replace $V \to e^{i2\pi N_{1}} = e^{i\phi_{{1}}(0)}$.
Remembering that at the fixed point the two leads are effectively well-coupled, $N_{1}$
behaves as the population of one half of an
infinite lead. With this, the correlation function of $V$ with itself
decays in time as $1/t$, again leading to a logarithmic divergence of
the susceptibility with the largest cutoff energy (magnetic field,
temperature, or frequency).

\subsection{Spin-asymmetric Rabi-Kondo}

We now add to the Kondo effect a laser, which tries to Rabi-flip the
electron constituting the impurity spin into a level decoupled
from the leads. We will introduce a corresponding two-level degree of
freedom, with Pauli matrices $\tau_{x,y,z}$, whose two states
$\tau_z = \pm 1$ correspond to the electron in the coupled conduction ($c$) level
(Kondo) and in the valence ($v$) level (trion), 
respectively. The Rabi flopping ($\tau_x$) is induced by a laser with
amplitude $\Omega$. If the laser has a proper circular polarization,
it only couples to a spin-up electron, $S_z = 1/2$.

\subsubsection{Single-channel spin-asymmetric Rabi-Kondo} 

Let us start from the single-channel spin-asymmetric Rabi-Kondo (1CARK)
case, analyzed in our previous work~\cite{Sbierski2013}. Based on all
the above considerations, the Hamiltonian is:
\begin{equation} 
\begin{aligned}
H_\onecark &= \frac{u}{4\pi}
  \int_{-\infty}^\infty \mathrm{d}x \left[ \partial_x \phi(x)
  \right]^2 + \frac{J_z}{\pi\sqrt{2}} \PK 
  S_z \partial_x \phi(0) \\
& \quad + \frac{J_{xy}}{2\pi a} \PK 
  \left(S_+ e^{i\sqrt{2}\phi(0)} + \mathrm{h.c.} \right) + 2 \Omega
  \tau_x P_\uparrow .
\end{aligned}
\label{eqn:h1cark}
\end{equation}
Here $\PK = \tfrac{1}{2}(1 + \tau_z)$ acts as
a local projector onto the $c$ level (i.e., the Kondo sector), and
$P_\uparrow = \tfrac{1}{2} + S_z$ as a local projector onto the
spin-up subspace.  We will concentrate on the case where the Kondo
temperature is much larger than the Rabi frequency, $\Tk \gg \Omega$.
Then, at energy scales larger than $\Tk$, we can ignore the Rabi
term. The transformations and RG flow are as above, with the only
difference that every transformation
$U_\alpha = e^{-i \alpha S_z \phi(0)}$ should be replaced by
$U_\alpha = e^{-i \alpha (1+\tau_z) S_z \phi(0)/2}$. The series of
transformations on the way to the Kondo fixed point 
at $\mathcal{J}_z = 1$ then modifies the
Rabi term, giving
\begin{equation}
\begin{aligned}
H_\onecark^\mr{int} &= \frac{u}{4\pi} \int_{-\infty}^\infty \mathrm{d}x \left[ \partial_x \phi(x) \right]^2 + \frac{J_{xy}^\mathrm{ren}}{\pi a} \PK S_x \\
& \quad + \Omega P_\uparrow \left(\tau_+ e^{-i\phi(0)/\sqrt{2}} + \mathrm{h.c.}\right) ,
\end{aligned}
\label{eqn:hi1cark}
\end{equation} 
where $J_{xy}^\mathrm{ren} \propto \Tk$ $(\gg \Omega)$, as
mentioned above. Thus, the corresponding Kondo term is much larger
than the Rabi term, and effectively 
eliminates the $S_z$ part of $P_\uparrow = \tfrac{1}{2} + S_z$
(the eliminated part breaks the symmetry under $S_z \to -S_z$
on the scale $\Omega^*$ introduced below, as a
local magnetic field would do, but this has a negligible effect in the
current 1CK physics, since $\Omega^* \ll \Tk$). With this the Rabi
term looks exactly like the spin-flip exchange term in the pure Kondo
problem, Eq.~\eqref{eqn:hp1ck}, demonstrating that the Rabi term leads
to a secondary Kondo screening process.
The scaling dimension, say $\eta_1$, of the 1CRK term 
is dictated by the bosonic exponent, giving $\etaone = 1/4$,
reflecting the Anderson orthogonality catastrophe with a phase shift
change of $\pi/2$ in each spin channel caused by a Rabi flop. It can
also be thought of as a boundary condition changing operator (from
Kondo to non-Kondo), and CFT analysis~\cite{Affleck1994} gives the
same result for its scaling dimension. As a result, for frequencies in
the range $\Omega^* \ll |\omega| \ll \Tk$ (where the new low-energy
scale $\Omega^*$ will be defined shortly), the emission spectrum
(imaginary part of the retarded correlator of the Rabi term with
itself) scales as $|\omega|^{2\etaone-1} =
|\omega|^{-1/2}$. Moreover, the RG equation for $\Omega$ is~\cite{Cardy1996}
\begin{equation} \label{eqn:rgomega1c}
  -D \frac{\mathrm{d} (\Omega/D)}{\mathrm{d}D} = \left(1-\etaone \right) \frac{\Omega}{D},
\end{equation}
with solution $\Omega(D)/D = \Omega/\Tk (D/\Tk)^{\etaone -1}$,
where we have taken  into account
that the  RG flow of $\Omega$ starts at the scale of $\Tk$.
Therefore $\Omega(D)$ flows to strong coupling. The scale at
which $\Omega(D)/D$ becomes of order unity defines the renormalized Rabi
frequency (secondary Kondo temperature),
$\Omega^*/\Tk \sim (\Omega/\Tk)^{1/(1-\etaone)} = (\Omega/\Tk)^{4/3}$.
Thus, the impurity entropy starts with the value $\ln 3$ at $T \gg \Tk$ (the
four possible values of $S_z$ and $\tau_z$, except the excluded
possibility of $\tau_z=-1$ and $S_z=-1/2$), then decreases to $\ln 2$
for $\Omega^* \ll T \ll \Tk$ (due to the Kondo screening of the
$\tau_z=1$ sector), and then goes to zero for $T \ll \Omega^*$, due to
the secondary Kondo screening.

Below $\Omega^*$, secondary Kondo screening (of the $\tau$ degree of
freedom) sets in. The emission spectrum, which corresponds to a
correlator of the Rabi term with itself, becomes the spectral function
of the correlator of the secondary Kondo exchange term with
itself. Our previous analysis for the single-channel case shows that
this leads to an $|\omega|^3$ behavior, or, in the presence of detuning
(which adds to the Hamiltonian a term proportional to $\tau_z$, that
is, a magnetic field in the secondary Kondo language), to an $|\omega|$
scaling. Also, at zero frequency a delta function appears in the
emission spectrum. Its amplitude can be calculated in two ways. One
is to note that the spectral weight missing by the emergence of
$\Omega$ and the corresponding change of the spectral function from
$\propto |\omega|^{2 \etaone -1}$
to a positive power should go into the delta
function, giving it a weight scaling as
$\int_0^{\Omega^\ast} \!\! d \omega \, |\omega|^{2 \etaone - 1} \sim
\Omega^{\ast 2 \etaone} \sim \Omega^{2 \etaone/(1-\etaone)} =
\Omega^{2/3}$.
The other argument is that the coefficient of the delta function is
$|\langle \textrm{G}| \tau_x | \textrm{G}\rangle|^2$, the square
of the matrix element of $\tau_x$ (before the transformations) between
the ground state and itself, and this matrix element 
is the ground-state expectation value of the Rabi term divided by
$\Omega$. The expectation value of the Rabi term 
scales as $\Omega^*$, giving again an $(\Omega^\ast/\Omega)^2 =
\Omega^{2\etaone/(1-\etaone)} = \Omega^{2/3}$ scaling of the weight
of the delta function.

\begin{widetext}
\subsubsection{Two-channel spin-asymmetric Rabi-Kondo}

We will now consider the analogous two-channel spin-asymmetric   Rabi-Kondo (2CARK)  setup. Now the starting Hamiltonian is:
\begin{equation} \label{eqn:h2cark}
H_\twocark = \sum_{\ell=1,2}
\left\{ \frac{u}{4\pi} \int_{-\infty}^\infty \mathrm{d}x \left[ \partial_x \phi_\ell(x) \right]^2 + \frac{J_z}{\pi\sqrt{2}} 
\PK S_z \partial_x \phi_\ell(0) + \frac{J_{xy}}{2\pi a} \PK
\bigl( S_+ e^{i\sqrt{2}\phi_\ell(0)} + \mathrm{H.c.} \bigr) \right\} + 2 \Omega 
P_\uparrow \tau_x.
\end{equation}
At energies larger than $\Tk$, we can use similar steps to the above, and arrive at:
\begin{equation} \label{eqn:hi2cark}
H_\twocark^\mathrm{int} = \sum_{p=\pm} \frac{u}{4\pi} \int_{-\infty}^\infty \mathrm{d}x \left[ \partial_x \phi_\ell(x) \right]^2 + \frac{J_{xy}^\mathrm{ren}}{\pi a} \PK
S_x \cos \left[\phi_-(0)\right] + \Omega P_\uparrow
\bigl(\tau_+ e^{-i\phi_+(0)/2} + \mathrm{h.c.}\bigr)
\end{equation}  
\end{widetext}

For the 2CRK model, the scaling dimension, say $\etatwo$, of the Rabi
term, seen as a boundary condition changing operator, is given by
$\etatwo=3/16$~\cite{Affleck1994}. One could arrive at this value
using also our abelian bosonization language: The $\phi_+$ exponent
contributes 1/8 to the scaling dimension of the Rabi term. Beyond
that, the $\tau_\pm$ operators turn on or off the transformed Kondo
exchange term involving $\cos[\phi_-(0)]$.
Now, turning on and off such a cosine appears in the problem of the Fermi edge
singularity, that is, turning on and off backscattering by impurity,
in a Luttinger liquid. This problem was analyzed in
Refs.~\cite{Gogolin1993,Prokofev1994}. They showed that, at long
times, the cosine can be replaced by a quadratic term (since it is
relevant), which allows one to find its contribution to the long time
behavior of the correlation function of $\tau_x$. This contribution
scales as $t^{-1/8}$, corresponding to a scaling dimension of
1/16. Adding this to the $1/8$ contributed by the exponential of $\phi_+$,
we recover the CFT
result $\etatwo=3/16$.  Thus, for $\Omega^* \ll |\omega| \ll \Tk$ the
emission spectrum behaves as $|\omega|^{2\etatwo-1} =
|\omega|^{-5/8}$.
The RG equation for $\Omega$ is the same as above, 
with $\etatwo$ taking the place of $\etaone$, 
reflecting the different scaling dimension of the Rabi term.
Then we get a low-enegry scale
$\Omega^* \propto \Omega^{1/(1-\etatwo)} = \Omega^{16/13}$, and the
weight of the delta peak at zero frequency scales as
$(\Omega^*)^{2\etatwo} \propto \Omega^{2\etatwo/(1-\etatwo)} =
\Omega^{6/13}$.
The impurity entropy will be $\ln 3$ for $T \gg \Tk$,
$\ln(1+\sqrt{2})$ (2CK partial screening + exciton state) for
$\Omega^* \ll T \ll \Tk$, and zero for $T \ll \Tk$.

As for the behavior of the emission spectrum at
$|\omega| \ll \Omega^*$, one could argue that the Rabi term, with its
explicit $S_z$ dependence, breaks the symmetry for flipping $S_z$ and
has similar effects to a local magnetic field on the physical
spin. Thus, below $\Omega^*$ the 2CK physics should be suppressed, and
one should recover the 1CK behavior of $|\omega|^3$ or $|\omega|$ in the
absence or presence of detuning, respectively.

\subsection{Spin-symmetric Rabi-Kondo}

Finally we arrive at the spin-symmetric version of the Rabi-Kondo
problem, where the applied laser features the two circular
polarizations with the same amplitude (i.e., a linear polarization),
and thus couples equally to both spin states.

\subsubsection{Single-channel spin-symmetric Rabi-Kondo}

We start from the single-channel spin-symmetric Rabi-Kondo (1CSRK)
problem. Now the Hamiltonian is:
\begin{equation}
\begin{aligned}
H_\onecsrk &= \frac{u}{4\pi} \int_{-\infty}^\infty \mathrm{d}x \left[ \partial_x \phi(x) \right]^2 + \frac{J_z}{\pi\sqrt{2}} \PK S_z \partial_x \phi(0) \\
& + \frac{J_{xy}}{2\pi a} \PK
\left(S_+ e^{i\sqrt{2}\phi(0)} + \mathrm{h.c.}\right)  + 2 \Omega \tau_x.
\end{aligned}
\label{eqn:h1csrk}
\end{equation}
Here on the scale of $\Tk$ we obtain:
\begin{equation}
\begin{aligned}
H_\onecsrk^\mr{int} &= \frac{u}{4\pi} \int_{-\infty}^\infty \mathrm{d}x \left[ \partial_x \phi(x) \right]^2 + \frac{J_{xy}^\mathrm{ren}}{\pi a} \PK
S_x \\
& \quad + \Omega \tau_+  \left[ P_\uparrow e^{-i\phi(0)/\sqrt{2}} + 
P_\downarrow e^{i\phi(0)/\sqrt{2}} \right]   + \mathrm{h.c.},
\end{aligned}
\label{eqn:hi1csrk}
\end{equation}
which is invariant under flipping of $S_z$, together with the lead (integrated) spin density $\phi(x)$. However, this symmetry is not essential in the 1CK case, and the analysis goes basically the same as in the spin-asymmetric case (at least as long as one considers the spin-symmetric emission spectrum).

Let us note that one could formally map the secondary screening
problem to an anisotropic spin-1 Kondo problem. Indeed, since
$J_{xy}^\mathrm{ren}/(\pi a) \sim \Tk$ is large, we can
discard the $S_x=-1/2$ state in the basis of Eq.~\eqref{eqn:hi1csrk}, and be
left with three states: the Kondo state ($\tau_z=1$ and 
$S_x = -1/2$) and the two spin states of the exciton ($\tau_z =-1$ and
arbitrary spin). Introducing corresponding spin-1 operators, the
Hamiltonian would look like the single-channel spin-1 Kondo problem,
after the non-spin-flip term has been eliminated by a transformation
like those above, that modifies the exponents of the spin-flip
term. However, this implies that the secondary $J_z/(\pi u)$ is of
order 1, i.e., the spin-1 problem is strongly spin-anisotropic. Now,
any spin exchange anisotropy would cause the creation of impurity-spin
terms proportional to the square of the $z$ component of the effective
spin 1, which amounts to detuning the exciton and primary-Kondo
states. For weak anisotropy, it is sufficient to add a corresponding
compensating term to restore the degeneracy and hence the spin-1 Kondo
physics. However, in our case, where the bare secondary exchange
anistropy is very large, the physics never reaches the underscreened spin-1 Kondo regime.  Thus, the
impurity entropy goes from $\ln 4$ to $\ln 3$ and then to zero as $T$
is lowered through $\Tk$ and $\Omega^*$.

\begin{widetext}
\subsubsection{Two-channel spin-symmetric Rabi-Kondo}

The last case is the two-channel symmetric Rabi-Kondo (2CSRK) model,
with Hamiltonian
\begin{equation} \label{eqn:h2csrk}
H_\twocsrk = \sum_{\ell=1,2}
\left\{ \frac{u}{4\pi} \int_{-\infty}^\infty \mathrm{d}x \left[ \partial_x \phi_\ell(x) \right]^2 + 
\frac{J_z}{\pi\sqrt{2}} \PK 
S_z \partial_x \phi_\ell(0) + \frac{J_{xy}}{2\pi a} 
\PK \bigl(S_+ e^{i\sqrt{2}\phi_\ell(0)} + \mathrm{h.c.}\bigr) \right\} + 2 \Omega \tau_x,
\end{equation}
which becomes on the scale of $\Tk$:
\begin{equation} \label{eqn:hi2csrk}
H_\twocsrk^\mathrm{int} = \sum_{p=\pm} \frac{u}{4\pi} \int_{-\infty}^\infty \mathrm{d}x \left[ \partial_x \phi_\ell(x) \right]^2 +  
\frac{J_{xy}^\mathrm{ren}}{\pi a} \PK 
 S_x \cos \left[\phi_-(0)\right] + \Omega \tau_+ \left[ 
 P_\uparrow 
e^{-i\phi_+(0)/2} + P_\downarrow
e^{+i\phi_+(0)/2} \right]  + \mathrm{h.c.}
\end{equation}
\end{widetext}

Again the analysis parallels the spin-asymmetric case, except that now flipping $S_z$ together with  $\phi_+(x)$ 
remains a symmetry, so the 2CK physics is not destroyed at low energies, and a decoupled Majorana zero mode remains.
Indeed, while the Rabi term contains $S_z \propto S_x S_y$, the corresponding processes are suppressed by the dominant $J_{xy}^\mathrm{ren}$ term, and all higher order processes (in terms of $\Omega$) which leave the system within the low-energy manifold of $J_{xy}^\mathrm{ren}$ term do not couple to $S_y$.
It should show up in the correlation function of the Rabi term with itself, which depends on $S_z \propto S_x S_y$, and reduce one power of $\omega$ from the power-law dependence of the emission spectrum on $\omega$ for $|\omega| \ll \Tk$, that is, make it go as $|\omega|^2$ instead of $|\omega|^3$ (in the absence of detuning). Correspondingly, the impurity entropy goes from $\ln 4$ to $\ln(2+\sqrt{2})$ to $\ln \sqrt{2}$ as $T$ is decreased through $\Tk$ and $\Omega^*$.

%\bibliography{RabiKondo}

\end{document}